\documentclass[fleqn,usenatbib]{mnras}

\usepackage{newtxtext,newtxmath}

\usepackage[T1]{fontenc}

\DeclareRobustCommand{\VAN}[3]{#2}
\let\VANthebibliography\thebibliography
\def\thebibliography{\DeclareRobustCommand{\VAN}[3]{##3}\VANthebibliography}

\usepackage{graphicx}	
\usepackage{amsmath}	

\usepackage{amssymb}	

\usepackage{color}
\usepackage{hyperref}
\usepackage{subfigure}
\usepackage{float}
\usepackage{enumerate}
\usepackage{mathrsfs}
\usepackage{geometry}
\usepackage{booktabs}
\usepackage{multirow}
\usepackage{comment}
\usepackage{threeparttable}

\title[Determination of QPO properties: a case study]{Determination of QPO properties in the presence of strong broad-band noise: a case study on the data of MAXI J1820+070}

\author[D. K. Zhou et al.]{Deng-Ke Zhou$^{1,2}$, 
Shuang-Nan Zhang$^{1,2}$\thanks{Corresponding author. E-mail:
zhangsn@ihep.ac.cn},
Li-Ming Song$^{1,2}$\thanks{Corresponding author. E-mail: songlm@ihep.ac.cn},
Jin-Lu Qu$^{1,2}$,
Liang Zhang$^{1,3}$,
Xiang Ma$^{1}$,
\newauthor
You-Li Tuo$^{1}$,
Ming-Yu Ge$^{1}$,
Yanan Wang$^{3}$,
Shu Zhang$^{1}$ and
Lian Tao$^{1}$
\\
$^{1}$Key Laboratory of Particle Astrophysics, Institute of High Energy Physics, Chinese Academy of Sciences, Beijing, China\\
$^{2}$University of Chinese Academy of 
Sciences, Chinese Academy of Sciences, Beijing, China.\\
$^{3}$Physics \& Astronomy, University of Southampton, Southampton, Hampshire SO17~1BJ, UK
}

\date{Accepted XXX. Received YYY; in original form ZZZ}

\pubyear{2021}

\begin{document}
\label{firstpage}
\pagerange{\pageref{firstpage}--\pageref{lastpage}}
\maketitle

\begin{abstract}
Accurate calculation of the phase lags of quasi-periodic oscillations (QPOs) will provide insight into their origin. In this paper we investigate the phase lag correction method which has been applied to calculate the intrinsic phase lags of the QPOs in MAXI J1820+070. We find that the traditional additive model between BBN and QPOs in the time domain is rejected, but the convolution model is accepted. By introducing a convolution mechanism in the time domain, the Fourier cross-spectrum analysis shows that the phase lags between QPOs components in different energy bands will have a simple linear relationship with the phase lags between the total signals, so that the intrinsic phase lags of the QPOs can be obtained by linear correction. The power density spectrum (PDS) thus requires a multiplicative model to interpret the data. We briefly discuss a physical scenario for interpreting the convolution. In this scenario, the corona acts as a low-pass filter, the Green's function containing the noise is convolved with the QPOs to form the low-frequency part of the PDS, while the high-frequency part requires an additive component. We use a multiplicative PDS model to fit the data observed by \emph{Insight}-HXMT. The overall fitting results are similar compared to the traditional additive PDS model. Neither the width nor the centroid frequency of the QPOs obtained from each of the two PDS models were significantly different, except for the r.m.s. of the QPOs. Our work thus provides a new perspective on the coupling of noise and QPOs.
\end{abstract}



\section{Introduction}
Decades of research on black hole binaries (BHBs) show that their X-ray emission is variable on different time scales, including the low-frequency (mHz to 30Hz) quasi-periodic oscillations (QPOs) and the broad-band noise (BBN) (\citealt{1999ApJ...520..262P,RN972,RN970,2016AN....337..398M}). The study of the timing signals can effectively diagnose the geometric characteristics of the disk and the corona near the black hole (\citealt{RN324,RN1009,RN958,RN802}). The disk and the corona near the black hole continuously radiate X-ray photons outward due to various radiation mechanisms (thermal radiation, Compton Radiation and so on). Photons with different energy arrive at the observer at different times because they may come from different radiation regions (\citealt{2000ApJ...531..963L,RN954}), or undergo different scattering processes (\citealt{1999ApJ...524L..59C, 2001AIPC..599..310P}), or have very complex mechanisms that cause delays (\citealt{1997ApJ...482..993M,1999ApJ...526L..33W,2010ApJ...710..836Q}). Therefore, analyzing the phase/time lags of photons between different energy bands helps us better understand the geometric or radiometric characteristics of X-ray BHBs. A common analysis method is based on Fourier cross-spectrum, which measures the frequency-dependent phase lag spectrum (FDPLS) between the signals in two different energy bands (\citealt{RN662}). This method allows to study the phase lags of two signals as a function of Fourier frequencies. Thus the phase lags between different components of timing signals, which usually originate from different physical processes (\citealt{1995ApJ...452..710N,2007A&ARv..15....1D,RN970}), can be studied separately. For example, \cite{2020MNRAS.494.1375Z} conducted a systematic study on the phase lag of the type-C QPO and found that the phase lag behaviour of the sub-harmonic of the QPO is very similar to that of the QPO fundamental component but the second harmonic of the QPO shows a quit different phase lag behaviour. \cite{2011MNRAS.414L..60U} investigated the phase lag of the BBN components of GX-339 and found that the large lags can be explained by viscous propagation of mass accretion fluctuations in the disk.

The traditional way to obtain the phase lag of the QPO components is to assume that the other components contribute weakly to the lag in the QPO frequency range, and then directly treat the values in the QPO frequency range as the phase lag of the QPO components (e.g., \citealt{1997ApJ...482..993M,1999ApJ...526L..33W,RN1030,2020MNRAS.494.1375Z}). However, the coexistence of various components makes it difficult to calculate any of the individual component. In particular, when the BBN is sufficiently strong in the QPO frequency range, there is no reason to ignore the effect of the BBN on the measured QPO phase lag. Despite attempts by some authors to ameliorate this dilemma by fitting different components of the cross-spectrum (e.g., \citealt{RN1059}), there is no broad consensus on how to obtain the intrinsic phase lag of the QPO in the presence of strong BBN. Therefore, it is difficult to determine the intrinsic properties (including phase lag) of the QPO in the presence of strong BBN.

In a recent work (\citealt{RN998}, hereafter Ma21) the authors attempt to correct for the original phase lags, which gives a clear physical picture using the corrected phase lags. Ma21 investigated the behavior of the QPO phase lags in MAXI J1820+070 using \emph{Insight}-HXMT observations and proposed a method to obtain the intrinsic phase lag of the QPO. In their analysis of the phase lags, they find that by subtracting the phase lags below the QPO frequency range they can obtain consistent QPO phase lags as functions of photon energy for all observations and can explain the lag behavior through the precession of a compact jet above the black hole. On the data they used, the PDS shows that the BBN components are too strong compared to the QPO component to ignore the contribution to the phase lags in the QPO frequency range (see panel c of figure 1 in Ma21). If they do not correct the phase lags for the QPO, the phase lags obtained from the original FDPLS will be affected by the BBN and thus are not intrinsic phase lags of the QPO. Although Ma21 applied this method to obtain consistent results of the phase lags, the rationale for doing so was not explained in detail, so the plausibility of this correction method needs to be tested. On the data they analyzed, some of the observations obtained phase lags with little difference before and after the correction, but some of the phase lags changed significantly (even the sign is totally reversed) before and after the correction. Therefore, we believe it is necessary to investigate under what conditions the correction is effective and how the QPO component is related to the BBN component. The motivation of this paper is to explore the mechanism behind the correction method used by Ma21 and to investigate the QPO properties in the presence of strong BBN in conjunction with the results obtained by Ma21.

Since we want to obtain the properties of a certain component (in our case, the QPO), and what we observe is some kind of superposition of all components, we have to face the problem of how these components contribute to the total signal. In this paper, when we refer to the term signal, we are referring to the light curve or the underlying time series. The total signal is defined as the time series that we directly observe and the sub-signals are the sub-components such as QPO and BBN that make up the total signal. Traditionally, it is believed that the BBN and the QPO are additive in the time domain and that they are incoherent at any frequency, which is why the PDS is fitted by the sum of several Lorentzian functions. \cite{RN966} proposed a possible relationship between QPO and BBN, where the QPO component and the BBN component are multiplied in the time domain. In this case, a convolution model is required for the fitting of the PDS in the frequency domain. Another way in which the QPO component and the BBN component are combined into a total signal in the time domain is convolution, which is usually caused by the response of the QPO signal in the region where the BBN component is generated (a model similar to this mechanism can be found in \citealt{2010MNRAS.404..738C}). The calculation of the FDPLS involves the Fourier orthogonal decomposition of the signal, so it can be expected that if the sub-signals form the total signal in different ways, then the relationship between the FDPLS of the total signal and the FDPLS of the sub-signals must be different.

This paper is structured as follows: Section \ref{sec:2} analyzes the relationships between the FDPLS as well as the PDS of total signals and sub-signals. An algorithm to generate two signals satisfied specific PDS and FDPLS simultaneously is also proposed. Besides, one possible way of coupling the QPO component and the BBN component in the time domain is discussed. In Section {\ref{sec:3}}, based on the results of Ma21's analysis of MAXI J1820+070 on phase lags, we argue that the QPO component and the BBN component constitute the total signal by convolution in the time domain. Using the data of MAXI J1820+070, we fit the PDS in different energy bands using the multiplicative PDS model and the 
traditional additive PDS model, and compare their differences. In addition, we also performed some simulations to rule out the possibility that the total signal appears to be the sum of the sub-signals in the time domain. Section \ref{sec:4} discusses and summarizes the whole paper.

\section{theory and simulation}
\label{sec:2}
\subsection{phase lag relationship}
Suppose that the expressions of non-zero mean signals $r_1(t)$, $r_2(t)$, $q_1(t)$, $q_2(t)$ at frequency $f_0$ can be written as:
\begin{equation}
    \label{eq:sine}
    \begin{aligned}
    r_1(t) &= R_1\sin (2\pi f_0 t+\phi_{r_1})+c_{r_1},\\
    r_2(t) &= R_2\sin (2\pi f_0 t+\phi_{r_2})+c_{r_2},\\
    q_1(t) &= Q_1\sin (2\pi f_0 t+\phi_{q_1})+c_{q_1},\\
    q_2(t) &= Q_2\sin (2\pi f_0 t+\phi_{q_2})+c_{q_2},\\
    \end{aligned}
\end{equation}
where $c_{r_1}$, $c_{r_2}$, $c_{q_1}$ and $c_{q_2}$ are the mean values of the corresponding signals; $R_1$, $R_2$, $Q_1$, $Q_2$ and $\phi_{r_1}$, $\phi_{r_2}$, $\phi_{q_1}$, $\phi_{q_2}$ are the amplitudes and the initial phases of the corresponding signals, respectively. The frequency $f_0$ can take any non-negative value including $0$. When $0$ is taken, it indicates that this is a constant signal. If the total signal is the sum of the sub-signals in the time domain (hereafter this kind of total signal is called the additive signal), i.e. :
\begin{equation}
    \begin{aligned}
     s_1(t) &=  r_1(t)+q_1(t),\\
     s_2(t) &=  r_2(t)+q_2(t),\\
    \end{aligned}
\end{equation}
then the phase difference (i.e. phase lag) between $s_1(t)$ and $s_2(t)$ can be written as:
\begin{equation}
    \label{eq:add_phase}
    \begin{aligned}
       & \Delta\phi_{\rm add}(s_2,s_1;f_0) \\
       &=\phi_{s_2}-\phi_{s_1}\\
        &={\rm Arg} [R_2\cos (\phi_{r_2})+Q_2\cos (\phi_{q_2}),R_2\sin (\phi_{r_2})+Q_2\sin (\phi_{q_2})]\\
        &-{\rm Arg} [R_1\cos (\phi_{r_1})+Q_1\cos (\phi_{q_1}),R_1\sin (\phi_{r_1})+Q_1\sin (\phi_{q_1})].
    \end{aligned}
\end{equation}
Here we use $\rm Arg$ $[a,b]$ to denote the argument of the complex $a+ib$, where $i$ is the imaginary unit. It can be seen from equation (\ref{eq:add_phase}) that if the total signal is the additive signal, the phase lag between the total signals depends on the amplitude and initial phase of each sub-signal.

If the total signal is convoluted by the sub-signals (hereafter this kind of total signal is called the convolved signal), then the phase lag between the total signal and the phase lag between the sub-signals satisfies a linear relationship, the proof of which will be given below. Still assume that the sub-signals satisfy equation (\ref{eq:sine}), but at this time the total signals are equal to the convolution of the sub-signals:\
\begin{equation}
    \begin{aligned}
       s_1(t)&=r_1(t)\otimes q_1(t),\\
       s_2(t)&=r_2(t)\otimes q_2(t),
    \end{aligned}
\end{equation}
where the sign $\otimes$ represents the convolution operation. The Fourier transform of the convolution of two signals is equal to the multiplication of their respective Fourier transforms. We can obtain the cross-correlation function (CCF) of $s_1(t)$ and $s_2(t)$ in the frequency domain:
\begin{equation}
    \begin{aligned}
       {\rm CCF}(f)&=\frac{R_1R_2Q_1Q_2}{64\pi^2}e^{-i[\Delta\phi(r_2,r_1)+\Delta\phi(q_2,q_1)]}\delta^4(f-f_0),
    \end{aligned}
\end{equation}
where $\Delta\phi(r_2,r_1)=\phi_{r_2}-\phi_{r_1}$ and $\Delta\phi(q_2,q_1) = \phi_{q_2}-\phi_{q_1}$ are the phase lag of the sub-signals. The phase lag of the two total signals $s_1$ and $s_2$ can be obtained by taking the argument of their CCF:
\begin{equation}
    \label{eq:conv_phase}
    \begin{aligned}
       \Delta\phi_{\rm con}(s_2,s_1;f_0)&={\rm Arg}[{\rm CCF}(f)]=\Delta\phi(r_2,r_1)+\Delta\phi(q_2,q_1).
    \end{aligned}
\end{equation}
That is to say, if the total signal is the convolved signal, the phase lag of the total signals is equal to the sum of the phase lag of the sub-signals.

We also note that \cite{RN964} argued that the QPO component is multiplied together with the broad component to form the observed signal. We now consider the phase lag relationship between the total signal composed of single frequency sub-signals by multiplying them together (hereafter, this kind of total signal is called the multiplicative signal). $s_1(t)$ and $s_2(t)$ now are written as
\begin{equation}
    \label{eq:mul}
    \begin{aligned}
       s_1(t)&=r_1(t)\times q_1(t)\\&=c_{q_1}R_1\sin(2\pi f_0t+\phi_{r_1})+c_{r_1}Q_1\sin(2\phi f_0t+\phi_{q_1})\\
       &+\frac{1}{2}R_1Q_1\cos(2\pi\times2f_0t+\phi_{r_1}+\phi_{q_1})+c_{s_1},\\
       s_2(t)&=r_2(t)\times q_2(t)\\&=c_{q_2}R_2\sin(2\pi f_0t+\phi_{r_2})+c_{r_2}Q_2\sin(2\phi f_0t+\phi_{q_2})\\
       &+\frac{1}{2}R_2Q_2\cos(2\pi\times2f_0t+\phi_{r_2}+\phi_{q_2})+c_{s_2},
    \end{aligned}
\end{equation}
where $c_{s_1}$ and $c_{s_2}$ are constants. Thus, the two total signals $s_1(t)$ and $s_2(t)$ contain two non-zero frequency components, one at $f_0$ and the other at $2f_0$. We can see that the first two terms of $s_1(t)$ and $s_2(t)$ are in fact additive signals and thus the results on additive signals can be used. Thus, the phase lags of them can be written as:
\begin{equation}
    \label{eq:mul_phase}
    \begin{aligned}
    \Delta\phi_{\rm mul}(s_2, s_1;f_0)&=\Delta\phi_{\rm add}(s'_2,s'_1;f_0),\\
     \Delta\phi_{\rm mul}({s_2},{s_1};2f_0)&=\Delta\phi(r_2,r_1)+\Delta\phi(q_2,q_1),
    \end{aligned}
\end{equation}
where $s'_1=c_{q_1}R_1\sin(2\pi f_0t+\phi_{r_1})+c_{r_1}Q_1\sin(2\pi f_0t+\phi_{q_1})$ and $s'_2=c_{q_2}R_2\sin(2\pi f_0t+\phi_{r_2})+c_{r_2}Q_2\sin(2\pi f_0t+\phi_{q_2})$.
This is very interesting because the multiplicative signal seems to contain properties of both additive and convolved signals: on one hand the phase lag at frequency $f_0$ follows the pattern of the additive signal and on the other hand the phase lag at frequency $2f_0$ follows the pattern of the convolved signal. However, in general the mean value of the actual signal is larger than its amplitude, so it is expected that the total FDPLS of the multiplicative signal should be closer to the pattern of the 
additive signal, as we will see in the simulation section.

For the general signals $r_n$, $q_n$, $s_n$ ($n=0,1,... .N-1$), their discrete-time Fourier series are
\begin{equation} 
  \begin{split}
    r_n=\frac{1}{N}\sum_{k=0}^{N-1}R_ke^{i2\pi\frac{k}{N}n},\\
    q_n=\frac{1}{N}\sum_{k=0}^{N-1}Q_ke^{i2\pi\frac{k}{N}n},\\\
    s_n=\frac{1}{N}\sum_{k=0}^{N-1}S_ke^{i2\pi\frac{k}{N}n},
  \end{split}
\end{equation}
where $R_k$, $Q_k$, and $S_k$ are the discrete Fourier transforms of $r_n$, $q_n$, and $s_n$, respectively. Thus $r_n$, $q_n$, $s_n$ can be treated as a superposition of many trigonometric functions with different amplitude, different frequencies and different initial phases. For the additive and convolved signals discuss above, at the specified frequency, these signals have the same properties as the corresponding single frequency signals. So the phase lags of the additive signal follows equation (\ref{eq:add_phase}) at each frequency, and the phase lags of the convolved signal follows equation (\ref{eq:conv_phase}) at each frequency. In summary, for the additive/convolved signals, their FDPLS follow equation (\ref{eq:add_phase}) or (\ref{eq:conv_phase}) at each specific frequency, respectively. In this case, each frequency corresponds to a set of parameters (amplitude and initial phase) for calculating the phase lag. Unfortunately, it is clear from equation (\ref{eq:mul}) that additional frequency components appear in the multiplicative signal that are not identical to the sub-signals, so the conclusion for the single-frequency signal cannot be generalized to the general signal. Nevertheless, we can use simulation (see sec \ref{sec:2.4}) to explore the phase lag relationship between the multiplicative signals. 

\subsection{PDS relationship}
\label{sec:2.2}
Assuming $s(t)$ is the additive signal, i.e. $s(t)=r(t)+q(t)$. Let $P_{\rm s}(f)$, $P_{\rm r}(f)$ and $P_{\rm q}(f)$ be the PDS of the signals $s(t)$, $r(t)$ and $q(t)$, respectively. Considering the Fourier transform is linear, one obtains:
\begin{equation}
    \begin{aligned}
     P_s(f)&=|\mathscr{F}(r+q)|^2\\
        &=|\mathscr{F}(r)+\mathscr{F}(q)|^2\\
        &=P_r(f)+P_q(f)+\mathscr{F}(r)^*\mathscr{F}(q)+\mathscr{F}(r)\mathscr{F}(q)^*.
    \end{aligned}
\end{equation}
The last two terms are actually the cross-spectrum of the signals $r(t)$ and $q(t)$. If $r(t)$ and $q(t)$ are incoherent at all frequencies, then their cross-spectrum will converge to zero after averaging many signal realizations. Therefore the above equation is simplified to 
\begin{equation}
    <P_{\rm s}(f)>=<P_r(f)>+<P_q(f)>,
\end{equation}
where the <> sign indicates the average of many realizations of the signals. This indicates that the PDS of the sum of two incoherent signals is equal to the sum of their respective PDS.

Assuming $s(t)$ is the convolved signal, i.e. $s(t)=r(t)\otimes q(t)$. The Fourier transform of the total signal $s(t)$ is equal to the multiplication of the Fourier transforms of the sub-signals $r(t)$, $q(t)$, i.e.
\begin{equation}
    \begin{aligned}
    &\mathscr{F}(s)=\mathscr{F}(r) \mathscr{F}(q),\\
    &<P_s(f)>=<P_r(f)><P_q(f)>.
    \end{aligned}
\end{equation}
That is, the PDS of the convolved signal is equal to the multiplication of the PDS of the corresponding sub-signals. It can be easily generalized that if $s(t)=r(t)\otimes q(t)+p(t)$, and $r(t)\otimes q(t)$ is incoherent with $p(t)$, their PDS will satisfy
\begin{equation}
    <P_{\rm s}(f)>=<P_{\rm r}(f)><P_{\rm q}(f)>+<P_{\rm p}(f)>.
\end{equation}

The PDS properties for convolved signals can be generalized to multiplicative signals simply based on the symmetry of the Fourier transform, that is, the PDS of the multiplicative signal is the convolution of the PDS of the sub-signals.
\subsection{An algorithm for simultaneously simulating signals with specified PDS and FDPLS}
\label{sec:2.3}
In order to verify the correctness of the above theoretical analysis as well as to facilitate the analysis below, some simulations need to be done. An algorithm is thus needed to generate two signals with specified PDS and specified FDPLS simultaneously. The algorithm steps are as follows:
\begin{enumerate}[1)]
\item Use \cite{RN1007} (TK95 in the following) algorithm to generate two signals $s(t)$ and $s^{\prime}(t)$ that satisfy the specified PDS. Because the phase given to the signal by TK95 algorithm is random, the phase lag between these two signals is now on average zero. Denote their Fourier transforms as $S(f)$, $S^{\prime}(f)$, respectively.
\item Given the FDPLS $\phi(f)$, then calculate ${\rm CCF}(f)$ according to the following equation:
\begin{equation}
    {\rm CCF}(f)=\begin{cases}
    |S(f)||S^{\prime}(f)|\{\cos [\phi(f)]+i\sin [\phi(f)]\}&\text{$f\neq0$},\\
    |S(f)||S^{\prime}(f)|&
    \text{$f=0$},
    \end{cases}
\end{equation}
where $i$ is the imaginary unit.
\item The complex array ${\rm CCF}(f)$ obtained in step 2 is divided by the complex conjugate of $S(f)$ to obtain a new complex array. Then performing inverse Fourier transform to it to obtain the signal $s^{\prime\prime}(t)$. Expressed in mathematical notation, it is
\begin{equation}
    s^{\prime\prime}(t)=\mathscr{F}^{-1}[\frac{{\rm CCF}(f)}{S^*(f)}].
\end{equation}
\end{enumerate}

The underlying PDS of $s^{\prime\prime}(t)$ is the same as the PDS of $s(t)$, but the FDPLS between $s^{\prime\prime}(t)$ and $s(t)$ will satisfy the given FDPLS. In summary, $s(t)$ and $s^{\prime\prime}(t)$ satisfy both the given PDS and the given FDPLS. In this paper, all PDS and FDPLS are extracted using the X-ray astronomy python package \textit{stingray} (\citealt{2019ApJ...881...39H}, version 0.3), and all PDS and FDPLS fitting are done by \textit{XSPEC} (\citealt{1996ASPC..101...17A}, version 12.11.1) or \textit{lmfit} (\citealt{newville_matthew_2014_11813}, version 1.0.2).

\subsection{simulation}
\label{sec:2.4}
\begin{table*}
  \caption{Timing properties of $r_1(t)$, $r_2(t)$, $q_1(t)$, $q_2(t)$ (see section \ref{sec:2.4} for the definition of them).}
    \begin{tabular}{ccccccccc}
    \toprule
    signals & bin size (s) & $\nu_c$ & $\omega$ & mean rate (cts/s) & exposure (s) & fractional r.m.s & PDS type & FDPLS type \\
    \midrule
    $r_1(t)$ & 0.01 & 0 & 3 & 2000 & 2000 & 30\% & BBN & \multirow{2}[0]{*}{constant} \\
    $r_2(t)$ & 0.01 & 0 & 4 & 2000 & 2000 & 20\% & BBN &  \\
    \hline
    $q_1(t)$ & 0.01 & 1 & 0.1 & 2000 & 2000 & 15\% & QPO & \multirow{2}[0]{*}{dip} \\
    $q_2(t)$ & 0.01 & 1 & 0.2 & 2000 & 2000 & 10\% & QPO & \\
    \bottomrule
    \end{tabular}
  \label{tab:1}
\end{table*}
\begin{figure*}
    \centering
    \includegraphics[scale=0.5]{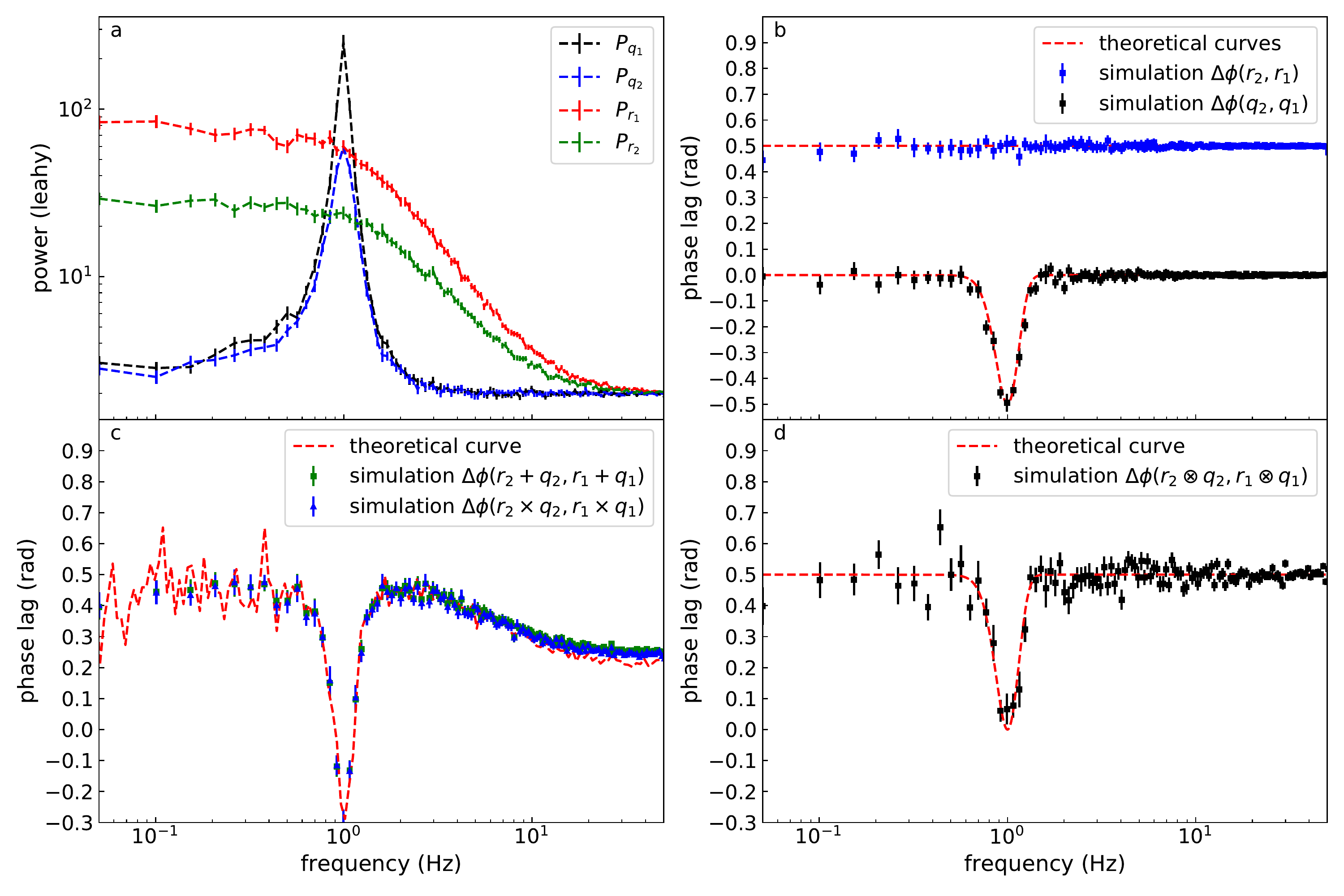}
    \caption{Simulation results of the FDPLS and PDS. Panel a: PDS of signals $r_1(t)$, $r_2(t)$, $q_1(t)$, $q_2(t)$. Panel b: simulated FDPLS (blue dots) between $r_1(t)$ and $r_2(t)$ and simulated FDPLS (black dots) between $q_1(t)$ and $q_2(t)$. Panel c: simulated FDPLS (green dots) between $r_2(t)+q_2(t)$ and $r_1(t)+q_1(t)$ and simulated FDPLS (blue dots) between $r_2(t)\times q_2(t)$ and $r_1(t)\times r_2(t)$. Panel d: simulated FDPLS between $r_2(t)\otimes q_2(t)$ and $r_1(t)\otimes q_1(t)$. The red curves in panels b, c, and d are theoretically calculated curves. Error bars correspond to $1\sigma$ confidence intervals.}
    \label{fig:simu_FDPLS}
\end{figure*}
Four signals $r_1(t)$, $r_2(t)$, $q_1(t)$, and $q_2(t)$ with time resolution of 0.01 s are simulated according to the algorithm proposed in subsection \ref{sec:2.3}. The PDS of all these signals are characterized by the Lorentzian function, which takes the form of
\begin{equation}
    L(f)=\frac{K(\omega/(2\pi))}{(\omega/2)^2+(f-f_c)^2},
\end{equation}
where $K$, $\omega$ and $f_c$ denote the normalization factor, the full width at half maximum (FWHM) and the centroid frequency, respectively. The PDS of $r_1(t)$ and $r_2(t)$ are modeled by setting the centroid frequency of the Lorentzian function to zero and taking a large $\omega$, which simulates BBN, while $q_1(t)$ and $q_2(t)$ are modeled by taking the appropriate non-zero centroid frequency and $\omega$, which simulates QPO. In addition, the theoretical FDPLS $\phi(f)$ is also set. The FDPLS between BBNs is set to be constant, while the FDPLS between QPOs is set to have a dip-like feature near the centroid frequency (as seen in MAXI J1820+070). That is,
\begin{equation}
    \phi(f)=\begin{cases}
    0.5&\text{for BBN},\\
    -0.5e^{\frac{(f-1)^2}{0.04}}&
    \text{for QPO}.
    \end{cases}
\end{equation}
The timing properties of these four signals are summarized in table \ref{tab:1}.
We then split each signal into multiple 20-sec segments and calculated the PDS of each segment with Leahy normalization (\citealt{RN332}). The PDS is rebined by a logarithmic factor of 0.03 and we finally obtain the averaged PDS with the frequency range of 0.05-50.53 Hz. The FDPLS is obtained using cross-spectrum analysis. 

\begin{figure*}
    \centering
    \includegraphics[scale=0.38]{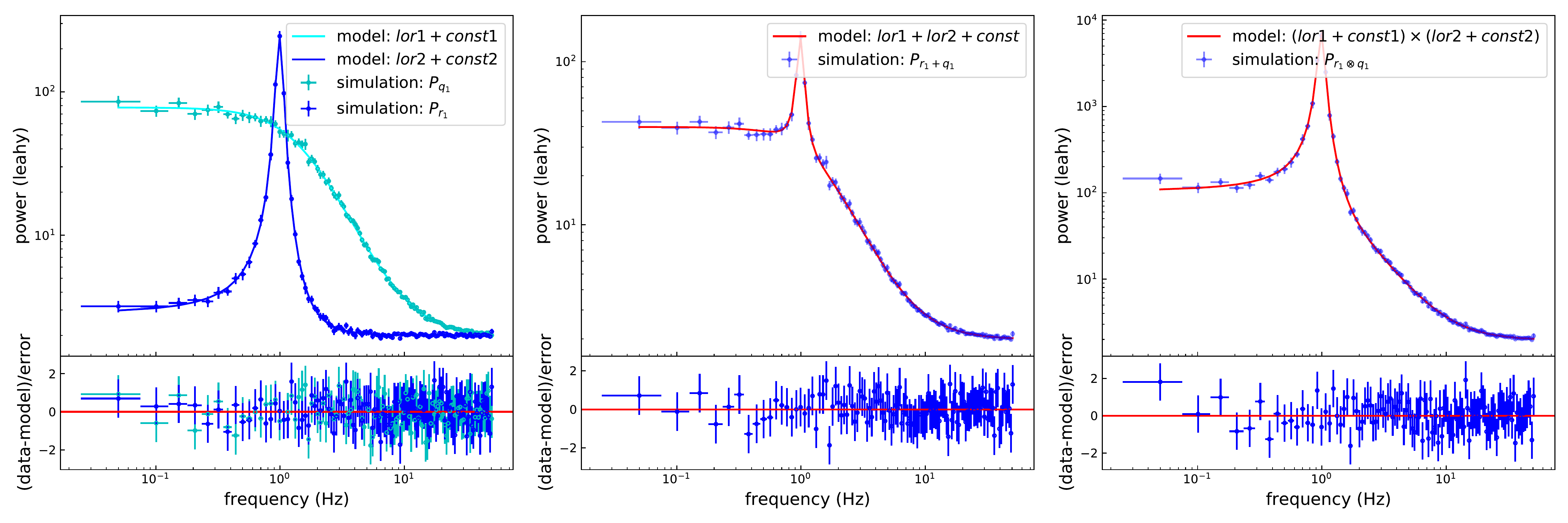}
    \caption{PDS of simulation results. Left panel: the cyan and blue data points are the PDS of the signals $r_1(t)$ and $q_1(t)$, respectively. The solid lines are the best-fit using Lorentzian model (considering the contribution of Poisson noise requires adding a constant to the Lorentzian model). Middle panel: the blue points are the PDS of the sum of the signals $r_1(t)$ and $q_1(t)$. The red solid line is the best-fit using two summed Lorentzian functions. Right panel: the blue points are the PDS of the convolution of the signals $r_1(t)$ and $q_1(t)$. The red solid line is the best-fit using two multiplicative Lorentzian functions (considering the contribution of Poisson noise requires adding a constant to each Lorentzian model). Error bars correspond to $1\sigma$ confidence intervals.}
    \label{fig:fitpds}
\end{figure*}

The results of the simulated PDS and the FDPLS are shown in panel a and panel b of Fig \ref{fig:simu_FDPLS}, respectively. When the total signal is assumed to be additive or multiplicative signal, the FDPLS between the total signals is shown in panel c of Fig.~\ref{fig:simu_FDPLS}. When the total signal is assumed to be the convolved signal, the FDPLS between the total signals is shown in panel d of Fig.~\ref{fig:simu_FDPLS}. As stated in the theoretical analysis section, the FDPLS of the multiplicative signal is very close to the FDPLS of the additive signal (see the green data points and the blue data points in panel c of Fig.~\ref{fig:simu_FDPLS}). Due to the symmetry of the Fourier transform to convolution and multiplication, the PDS section will only compare the differences between convolved and additive signals. In panels b, c, d of Fig.~\ref{fig:simu_FDPLS}, the data points are obtained by simulation and the red dashed lines are obtained from our theoretical calculation (the theoretical curve drawn in panel c of Fig.~\ref{fig:simu_FDPLS} is for the additive signal, and we did not draw the theoretical curve for the multiplicative signal because of the analytical difficulties). The theoretical curve shown in panel c of Fig.~\ref{fig:simu_FDPLS} is calculated by using the value of the simulated data (i.e., amplitude, initial phase), and it appears to fluctuate around the data points, which is due to the randomness deliberately introduced by the simulation algorithm (see TK95 for detail). The difference between panel c and panel d of Fig.~\ref{fig:simu_FDPLS} is mainly due to the different dependence of the FDPLS on the different kinds of signals (additive, multiplicative, convolved signals) on each sub-signal. The FDPLS between the convolved signals depends only on the FDPLS between sub-signals, independent of the other properties of the sub-signals. This is not the case for the additive and multiplicative signals. So it can be seen from panel c that the FDPLS depends on the relative power of sub-signals, while in panel d the FDPLS does not depend on the shape of the PDS of the sub-signals. In conclusion, the simulation results of the FDPLS are consistent with the theoretical analysis. The results of the PDS simulation results are shown in Fig.~\ref{fig:fitpds}. 

The PDS of $r_1(t)$ and $q_1(t)$ are shown in the left panel of Fig.~\ref{fig:fitpds}, and the PDS of the additive and convolved signals are shown in the middle and right panels of Fig.~\ref{fig:fitpds}, respectively. We can see that the PDS of the additive signal is the sum of the PDS of the sub-signals, while the PDS of the convolved signal is the multiplication of the PDS of the sub-signals. The solid lines running through the data points in the PDS are the best-fit using the additive and multiplicative Lorentzian models for the additive signal and the convolved signal, respectively. Overall, the simulation results are in good agreement with those of the theoretical analysis.
\subsection{A possible mechanism for introducing a convolution mechanism in the time domain}
\label{sec:QPO&noise}
Assuming that the orbit of matter around a black hole is circular and Keplerian. The equation can be derived based on the conservation of mass and angular momentum (e.g. \citealt{RN958}), i.e.
\begin{equation}
    \frac{\partial\Sigma}{\partial t}=\frac{3}{R}\frac{\partial}{\partial R}[R^{\frac{1}{2}}\frac{\partial}{\partial R}(\nu\Sigma R)],
\end{equation}
where $\Sigma=\rho H$ is the surface density of the corona or disk, and $\nu$ is the kinematic viscosity. Assuming that the surface density at $t=0$ is $\Sigma(t=0,R)=\delta(R-R_0)$ and $\nu$ is a constant, we will get
\begin{equation}
    g(R,t)=\frac{m}{12\pi\nu t}(\frac{R}{R_0})^{-\frac{1}{4}}I_{\frac{1}{4}}(\frac{RR_0}{6\nu t})e^{-\frac{R_0^2+R^2}{12\nu t}},
\end{equation}
where $I_{\frac{1}{4}}$ is the modified Bessel function and $g(R,t)$ is called the Green's function of the system. Under the condition that the system is linear, the surface density of any initial fluctuation $q(t)$ at position $R=R_0$ is the convolution of that fluctuation with the Green's function, i.e., $\Sigma(R,t)=q(t)\otimes g(R,t)$ (\citealt{RN958}). Denoting the mass accretion rate as $\dot{M}(R,t)$, then the luminosity corresponding to such a accretion rate is $L(R,t)\propto \Dot{M}(R,t)\propto \Sigma(R,t)\propto q(t)\otimes g(R,t)$. If we check the region $R<<R_0$, then we will get $g(R,t)\propto t^{\frac{5}{4}}e^{-\frac{R_0^2}{12\nu t}}$. The PDS of such a damped exponential signal is a zero-centred Lorentzian function (\citealt{RN958}). By introducing two types of white noise, one associated with the Green's function and the other superimposed on the QPO signal, we assume that the observed signal is expressed in the time domain as $s(t)=g(t)\otimes wn_1\otimes [q(t)+wn_2]$. We have assumed that q(t) has the form of QPO. Considering that the white noise and QPO signals are incoherent, a PDS of the combined signal will has the form
\begin{equation}
    P(R,f)\propto P_{b1}(R,f)P_q(R,f)+P_{b2}(R,f),
    \label{eq:PDS}
\end{equation}
where $P_{b1}$ denotes the first zero-centred Lorentzian function (i.e., the BBN1 component), $P_q$ denotes the non-zero centred Lorentzian function (i.e., the QPO component) and $P_{b2}$ denotes the second zero-centred Lorentzian function (i.e., the BBN2 component). Note that the former term of the above summation is due to the fluctuation propagation in the form of QPO and the latter term is due to the fluctuation propagation in the form of white noise,  which dominates different frequency ranges (we will see this in section \ref{sec:3}).

Furthermore, it is worth noting that the above result is valid only when $R<<R_0$ and the assumptions about the white noise and QPO fluctuations are satisfied. The total observed luminosity is the integral of the differential luminosity over the entire corona after considering the emissivity (\citealt{RN970}), but the form is very complicated. Nonetheless, it is still worthwhile to start with a simple model to explain the data. For this reason, when fitting the PDS of the real data with the multiplicative PDS model in section \ref{sec:3}, only a form similar to equation (\ref{eq:PDS}) will be considered.
\begin{figure*}
    \centering
    \includegraphics[scale=0.13]{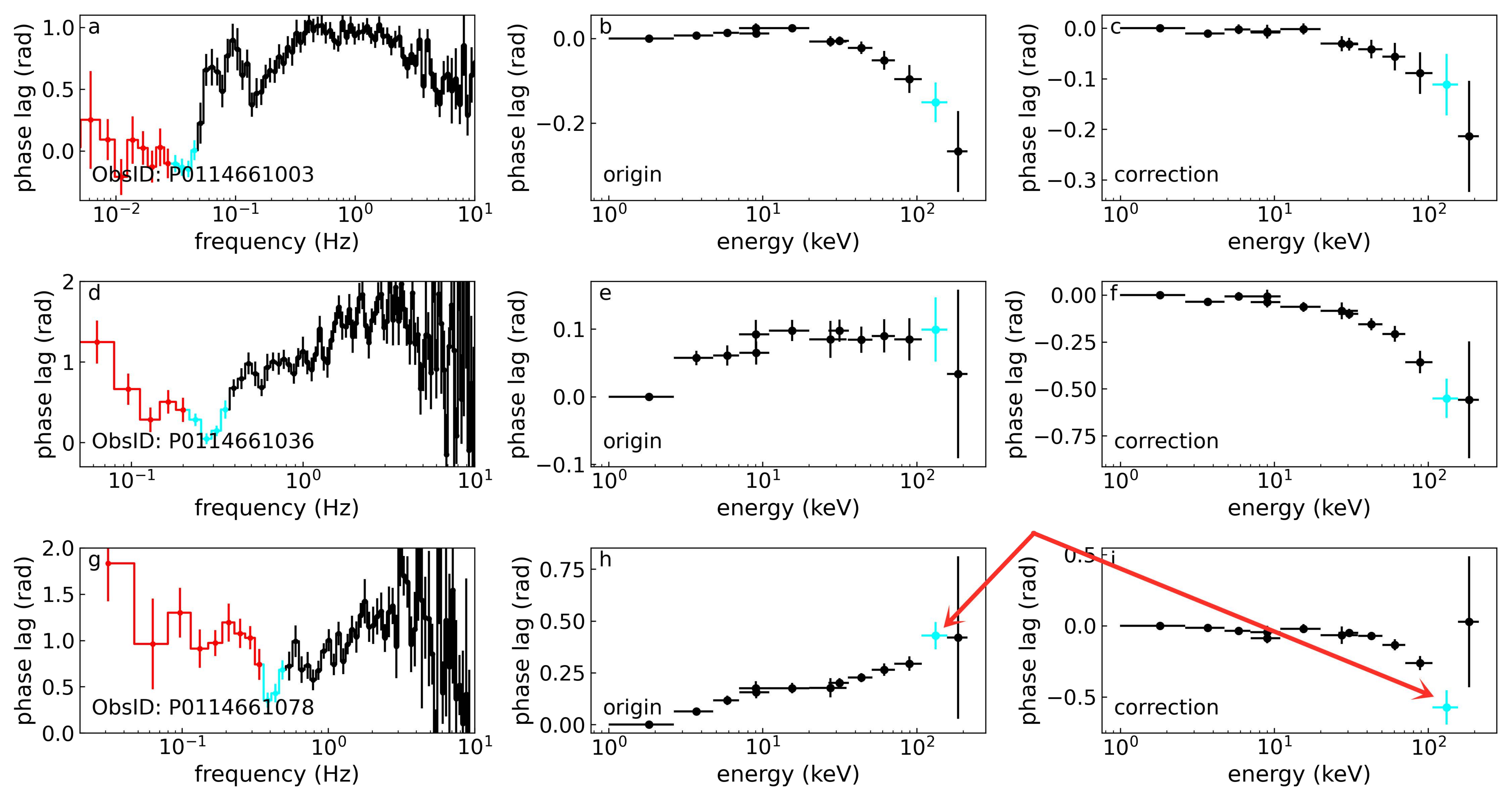}
    \caption{QPO phase lag correction of three typical \emph{Insight}-HXMT observations (reproduced from the data in Ma21). Each row represents the results of one observation. The FDPLS between 1-2.6 keV and 100-150 keV energy bands are shown in the left panels. Middle panels are the original QPO energy-dependent phase lags. Right panels are the intrinsic QPO energy-dependent phase lags after correction. The intrinsic QPO phase lags are obtained by subtracting the average of the phase lag of the BBN component (marked by the red dots on the left panels) from the original QPO phase lag (the averaged value marked by the cyan dots on left panels). The red arrows indicate the high energy band data that we will model in Fig.~\ref{fig:model_MAXI_FDPLS}. Error bars correspond to $1\sigma$ confidence intervals.}
    \label{fig:MAXI_FDPLS}
\end{figure*}
\begin{figure*}
	\centering
    \includegraphics[scale=0.16]{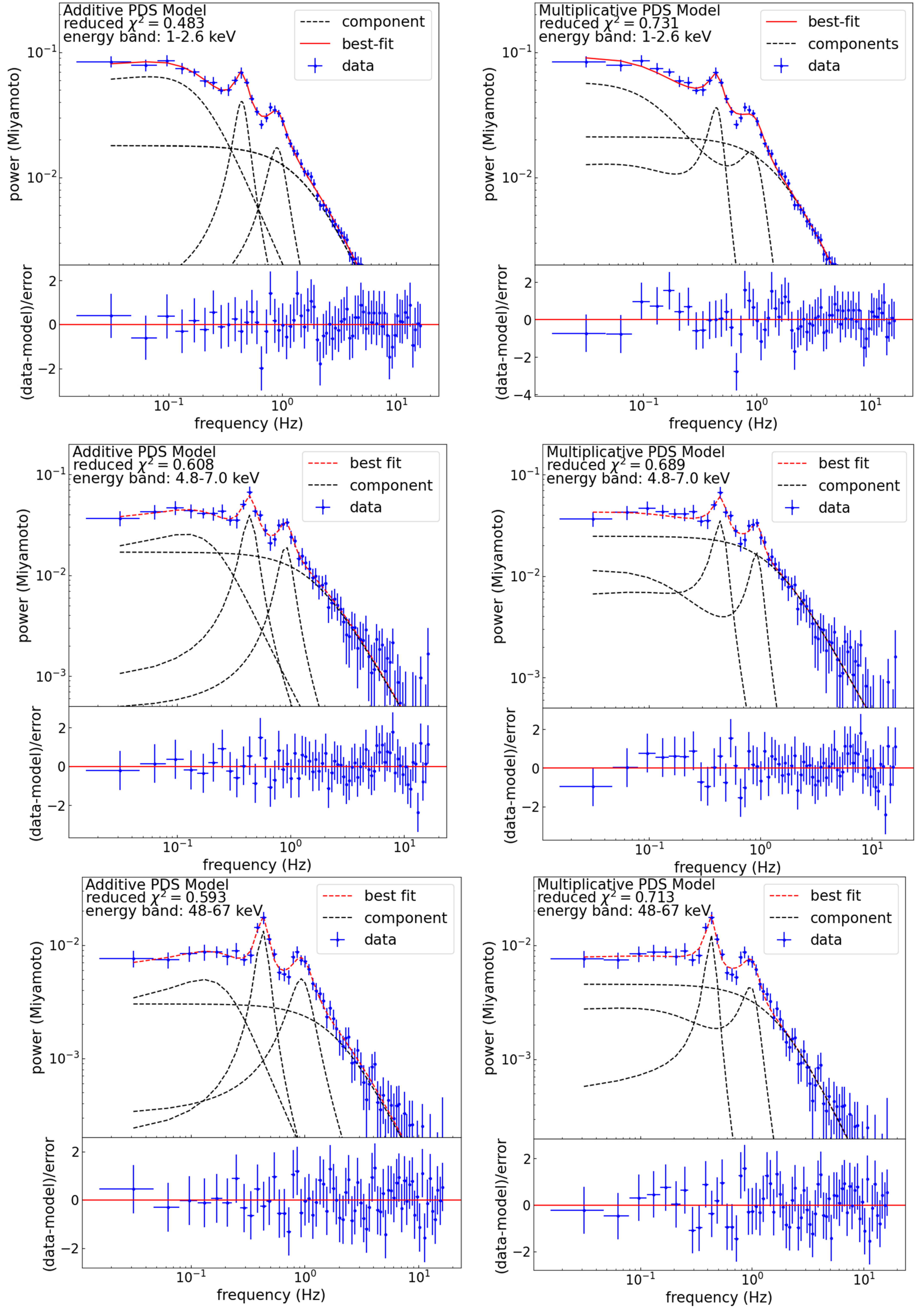}
	\caption{Several examples of the best-fit of the observed PDS data (ObsID P0114661078) in different energy bands with additive or multiplicative PDS models. The PDS model used in the left panels is the additive PDS model (i.e., equation (\ref{eq:ASM})). The PDS model used in the right panels is the multiplicative PDS model (i.e., equation (\ref{eq:MSM})). The contribution of Poisson noise in all PDS has been subtracted. Error bars correspond to $1\sigma$ confidence intervals.}
	\label{fig:ACPDS}
\end{figure*}
\begin{figure*}
    \centering
    \includegraphics[scale=0.3]{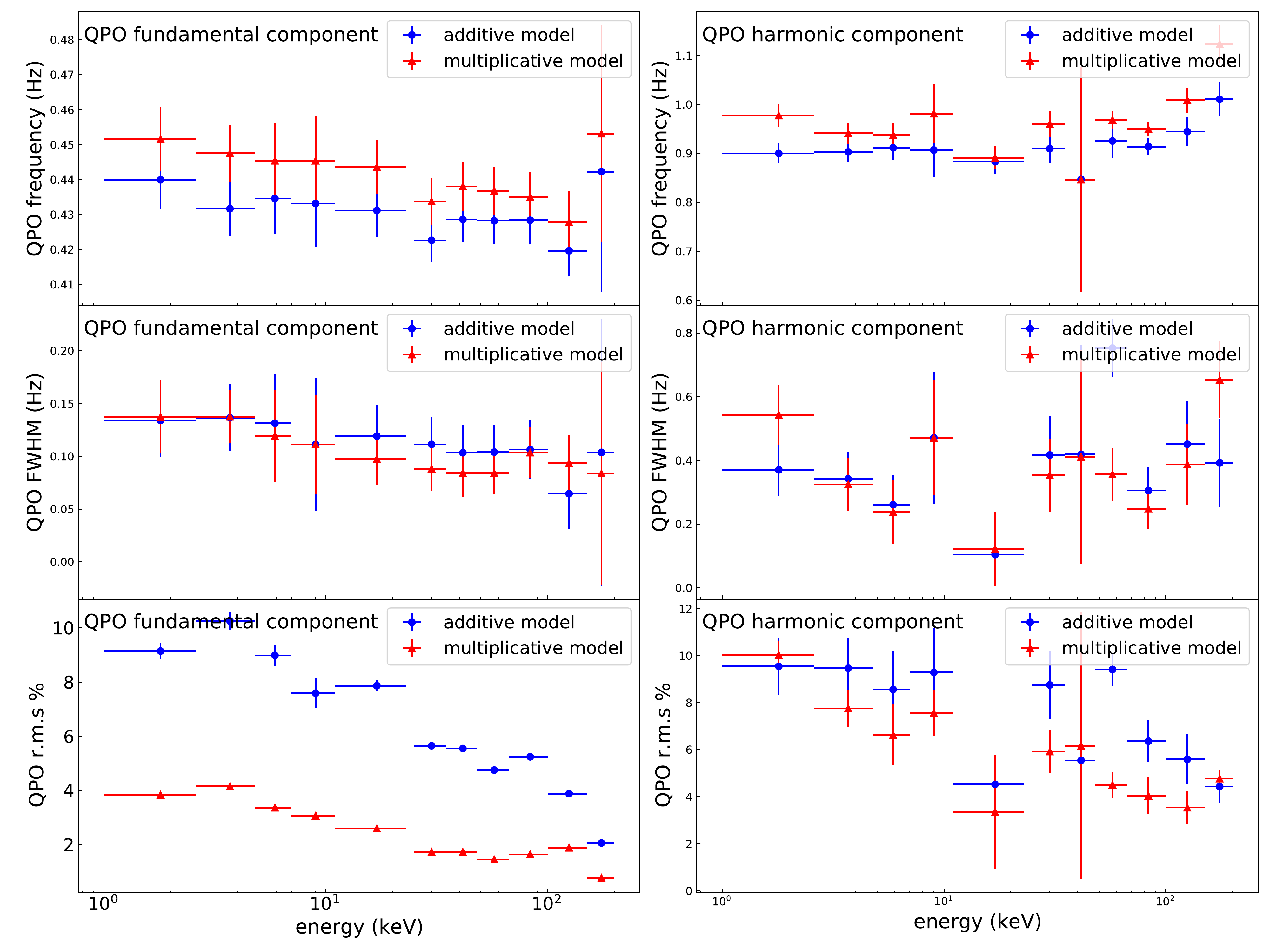}
    \caption{The parameters of the QPO (fundamental and harmonic components) as functions of photon energy which are obtained with the traditional additive and multiplicative PDS model, respectively. Error bars correspond to $1\sigma$ confidence intervals.}
    \label{fig:c_w_rms}
\end{figure*}
\section{The FDPLS and PDS of MAXI J1820+070}
\label{sec:3}
MAXI J1820+070 was discovered by the Monitor of All-sky X-ray Image (MAXI) during the outburst on 11 March 2018 (\citealt{2018ATel11399....1K}). It was confirmed to be a BHB (\citealt{Torres_2019}). \emph{Insight}-HXMT carried out observations three days after its discovery and obtained rich data with total exposure time of over 2000 ks. Ma21 has carried out a detailed temporal analysis of these data and, in particular, detailed calculations of the phase lags in different energy bands. Fig.~\ref{fig:MAXI_FDPLS} shows three typical observations from top to bottom, with a clear dip-like feature appearing near the QPO frequency range (the averaged value of this frequency range shown by the cyan dots denotes the original QPO phase lag). The phase lags of low-frequency BBN component are marked with red dots, denoted as background phase lag. The intrinsic phase lag of the QPO is obtained by subtracting the average of the background phase lag from the original QPO phase lag. After the correction, the absolute value of phase lag of the QPO becomes larger as the energy increases in all three observations. For the sake of clarity, the detailed correction steps used in Ma21 are re-summarized as follows:
\begin{enumerate}[1)]
    \item Calculate the FDPLS and identify the ccentroid frequency $f_{0}$ and the FWHM $\omega$ of the QPO according to the PDS.
    \item The original phase lag of the QPO is defined as the average of the phase lags in the frequency range $f_{0}\pm\omega/2$.
    \item The background phase lag is defined as the average of the phase lags below the QPO frequency range.
    \item The intrinsic phase lag of the QPO is then defined as the original phase lag minus the background phase lag.
\end{enumerate}
Such a correction actually implies two assumptions. The first assumption is that the phase lags of the BBN component at the QPO frequency range are the same as the phase lags below the QPO frequency range, at least their averaged values must be approximately equal. The second assumption is that the total phase lags (i.e., the observed original phase lags) at the QPO frequency range are equal to the sum of the BBN component phase lags and the QPO intrinsic phase lags. The correction of the phase lags is valid only when these two assumptions are satisfied simultaneously. The first assumption can be considered to be approximately satisfied. This is because the FDPLS obtained from MAXI J1820+070 shows that the phase lag does not vary significantly with frequency below the QPO frequency range. Thus it is reasonable to assume that at the QPO frequency range, the phase lags of the BBN component are approximately equal to the phase lags below the QPO frequency range. As to whether the second assumption can be satisfied, we need to first make an assumption about how the BBN component and the QPO component synthesize the observed signal. From the discussion of section \ref{sec:2} we know that if the observed signal is considered to be the sum of the BBN component and the QPO component (which is the default assumption in most of the literatures), the second condition cannot be satisfied. The second condition can be satisfied only when the observed signal is considered as a convolution of the BBN component and the QPO component. 

In the case that the observed signal is the convolved signal, the PDS of the signal needs to be fitted by a multiplicative PDS model. In section \ref{sec:QPO&noise} we introduced a multiplicative PDS model, so we use the multiplicative PDS model to fit the PDS in different energy bands and make a comparison with the traditional additive PDS model. We are not going to explore the multiplicative signal (i.e., the total signal is the multiplication of the sub-signals in the time domain) further because we are mainly concerned with the additive and convolved signals. For the additive PDS model, the form is
\begin{equation}
    \begin{split}
        P_{\rm add}(f)=&\frac{K_1(\omega_1/(2\pi))}{(\omega_1/2)^2+(f-f_{c_1})^2}+\frac{K_2(\omega_2/(2\pi))}{(\omega_2/2)^2+(f-f_{c_2})^2}\\&+\frac{K_3(\omega_3/(2\pi))}{(\omega_3/2)^2+(f-f_{c_3})^2}+\frac{K_4(\omega_4/(2\pi))}{(\omega_4/2)^2+(f-f_{c_4})^2}+c.
    \end{split}
    \label{eq:ASM}
\end{equation}
For the multiplicative PDS model, the form is
\begin{equation}
    \begin{split}
        P_{\rm mul}(f)=&\frac{K_1(\omega_1/(2\pi))}{(\omega_1/2)^2+(f-f_{c_1})^2}\times [\frac{K_2(\omega_2/(2\pi))}{(\omega_2/2)^2+(f-f_{c_2})^2}\\&+\frac{K_3(\omega_3/(2\pi))}{(\omega_3/2)^2+(f-f_{c_3})^2}]+\frac{K_4(\omega_4/(2\pi))^2}{(\omega_4/2)^2+(f-f_{c_4})^2}+c,
    \end{split}
    \label{eq:MSM}
\end{equation}
which is consistent with the multiplicative PDS model we discussed in section \ref{sec:QPO&noise}. We use these two models to fit the PDS in different energy bands in a representative \emph{Insight}-HXMT observation (ObsID P0114661078). The data reduction process in this paper is the same as Ma21. We extract the light curves with the time resolution of 0.03125 s in each energy band (1-2.6 keV, 2.6-4.8 keV, 4.8-7.0 keV, 7-11 keV, 11-23 keV, 25-35 keV, 35-48 keV, 48-67 keV, 67-100 keV, 100-150 keV and 150-200 keV). We then split the light curves into multiple 32-sec segments and calculate the PDS of each segment with Miyamoto normalization (\citealt{1991ApJ...383..784M}) for the convenience of calculating fractional r.m.s later, and finally obtain the averaged PDS with the frequency range of 1/32 to 16 Hz. After subtracting the contribution of Poisson noise in the PDS, we fitted the PDS with the additive and multiplicative PDS models. Some fitting examples in different energy bands are shown in Fig.~\ref{fig:ACPDS}. From top to bottom, Fig.~\ref{fig:ACPDS} shows the PDS fitting results for the three energy bands. The left panels are fitted using the traditional additive PDS model while the right panels are fitted using the multiplicative PDS model. In the multiplicative PDS model, the low-frequency zero-centered Lorentzian component is multiplied onto the QPO component instead of being added, resulting in the left side of the QPO component being lifted up in the right panels of Fig.~\ref{fig:ACPDS}. We can see that the total fitting results are similar but the individual components have some differences. Basing on the best-fit, we calculate the centroid frequency, the FWHM and the fractional r.m.s\footnote{The fractional r.m.s is calculated in the same way as for \cite{RN1033}, but ignoring the background correction, since the correction coefficients are the same for both models and our aim is only to compare their differences. Neglecting the background correction leads to a lower fractional r.m.s for the energy band with a lower signal-to-noise ratio (in this paper it is the higher energy band), but it does not change our conclusion.} of the QPO and parameters of the BBN on each energy band. And the results are listed in table \ref{tab:ACPDSfit}, \ref{tab:QPO_harmonic} and \ref{tab:BBN}. As shown in Fig.~\ref{fig:c_w_rms}, the centroid frequency and the FWHM of the QPO as functions of photon energy calculated according to the two models are similar, but the fractional r.m.s given by the two models are significantly different. For the fundamental component of QPO, the fractional r.m.s of QPO given by the traditional additive PDS model is about $2\sim 3$ times higher than that of the multiplicative PDS mode, but the trend is the same for both results. For the harmonic component of the QPO, the difference in the fractional r.m.s given by the two PDS models is not very significant.
\begin{table*}
  \centering
  \caption{Best-fit results for the fundamental frequency component of the QPO obtained using additive and multiplicative PDS models (i.e. equations (\ref{eq:ASM}) and (\ref{eq:MSM})), respectively.}
  \label{tab:ACPDSfit}%
  \begin{threeparttable}[b]
    \begin{tabular}{ccccc}
    \hline
       & \multicolumn{2}{c}{QPO frequency (Hz)} & \multicolumn{2}{c}{QPO FWHM (Hz)} \\
    \hline
    energy band (keV) & additive PDS model & multiplicative PDS model & additive PDS model & multiplicative PDS model \\
    \hline
    1.0-2.6 & $0.44\pm0.01$ & $0.45\pm0.01$ & $0.13\pm0.04$ & $0.14\pm0.03$ \\
    2.6-4.8 & $0.43\pm0.01$ & $0.45\pm0.01$ & $0.14\pm0.03$ & $0.14\pm0.03$ \\
    4.8-7.0 & $0.43\pm0.01$ & $0.45\pm0.01$ & $0.13\pm0.05$ & $0.12\pm0.04$ \\
    7.0-11.0 & $0.43\pm0.01$ & $0.45\pm0.01$ & $0.11\pm0.06$ & $0.11\pm0.05$ \\
    11.0-23.0 & $0.43\pm0.01$ & $0.44\pm0.01$ & $0.12\pm0.03$ & $0.10\pm0.03$ \\
    25.0-35.0 & $0.42\pm0.01$ & $0.43\pm0.01$ & $0.11\pm0.03$ & $0.09\pm0.02$ \\
    35.0-48.0 & $0.43\pm0.01$ & $0.44\pm0.01$ & $0.10\pm0.03$ & $0.08\pm0.02$ \\
    48.0-67.0 & $0.43\pm0.01$ & $0.44\pm0.01$ & $0.10\pm0.03$ & $0.08\pm0.02$ \\
    67.0-100.0 & $0.43\pm0.01$ & $0.44\pm0.01$ & $0.11\pm0.03$ & $0.10\pm0.02$ \\
    100.0-150.0 & $0.42\pm0.01$ & $0.43\pm0.01$ & $0.06\pm0.03$ & $0.09\pm0.03$ \\
    150.0-200.0 & $0.44\pm0.03$ & $0.45\pm0.03$ & $0.10\pm0.13$ & $0.08\pm0.11$ \\
    \hline
       & \multicolumn{2}{c}{QPO r.m.s \%\tnote{*}} & \multicolumn{2}{c}{reduced $\chi^2$} \\
    \hline
    energy band (keV) & additive PDS model & multiplicative PDS model & additive PDS model & multiplicative PDS model \\
    \hline
    1.0-2.6 & $9.14\pm0.31$ & $3.83\pm0.10$ & 0.48 & 0.71 \\
    2.6-4.8 & $10.26\pm0.31$ & $4.15\pm0.07$ & 0.74 & 0.88 \\
    4.8-7.0 & $8.99\pm0.40$ & $3.36\pm0.09$ & 0.61 & 0.69 \\
    7.0-11.0 & $7.59\pm0.56$ & $3.06\pm0.13$ & 0.66 & 0.66 \\
    11.0-23.0 & $7.86\pm0.20$ & $2.59\pm0.05$ & 0.48 & 0.57 \\
    25.0-35.0 & $5.65\pm0.11$ & $1.72\pm0.02$ & 1.02 & 1.55 \\
    35.0-48.0 & $5.54\pm0.13$ & $1.72\pm0.03$ & 0.89 & 1.14 \\
    48.0-67.0 & $4.75\pm0.09$ & $1.44\pm0.02$ & 0.59 & 0.71 \\
    67.0-100.0 & $5.24\pm0.13$ & $1.64\pm0.02$ & 0.45 & 0.58 \\
    100.0-150.0 & $3.88\pm0.14$ & $1.88\pm0.04$ & 0.43 & 0.44 \\
    150.0-200.0 & $2.05\pm0.12$ & $0.76\pm0.06$ & 0.44 & 0.41 \\
    \hline
    \end{tabular}%
    \begin{tablenotes}
     \item[*] No background correction is applied to the r.m.s because we are only interested in the difference between the results of fitting using the additive PDS model and the multiplicative PDS model, and the correction factors are the same for both models.
   \end{tablenotes}
  \end{threeparttable}
\end{table*}%

\begin{table*}
  \centering
  \caption{The same as table \ref{tab:ACPDSfit} but for the harmonic frequency component of the QPO.}
    \begin{tabular}{ccccc}
    \hline
      & \multicolumn{2}{c}{QPO frequency(Hz)} & \multicolumn{2}{c}{QPO FWHM(Hz)} \\
      \hline
    energy band (keV) & additive PDS model & multiplicative PDS model & additive PDS model & multiplicative PDS model \\
    \hline
    1.0-2.6 & $0.90\pm0.02$ & $0.98\pm0.02$ & $0.37\pm0.08$ & $0.54\pm0.09$ \\
    2.6-4.8 & $0.90\pm0.02$ & $0.94\pm0.02$ & $0.34\pm0.09$ & $0.33\pm0.08$ \\
    4.8-7.0 & $0.91\pm0.02$ & $0.94\pm0.02$ & $0.26\pm0.09$ & $0.24\pm0.10$ \\
    7.0-11.0 & $0.91\pm0.06$ & $0.98\pm0.06$ & $0.47\pm0.21$ & $0.47\pm0.18$ \\
    11.0-23.0 & $0.88\pm0.02$ & $0.89\pm0.02$ & $0.10\pm0.10$ & $0.12\pm0.12$ \\
    25.0-35.0 & $0.91\pm0.03$ & $0.96\pm0.03$ & $0.42\pm0.12$ & $0.35\pm0.11$ \\
    35.0-48.0 & $0.85\pm0.09$ & $0.85\pm0.23$ & $0.42\pm0.34$ & $0.41\pm0.34$ \\
    48.0-67.0 & $0.93\pm0.04$ & $0.97\pm0.02$ & $0.75\pm0.09$ & $0.36\pm0.08$ \\
    67.0-100.0 & $0.91\pm0.02$ & $0.95\pm0.02$ & $0.31\pm0.07$ & $0.25\pm0.06$ \\
    100.0-150.0 & $0.94\pm0.03$ & $1.01\pm0.03$ & $0.45\pm0.14$ & $0.39\pm0.13$ \\
    150.0-200.0 & $1.01\pm0.04$ & $1.12\pm0.04$ & $0.39\pm0.14$ & $0.65\pm0.12$ \\
    \hline
      & \multicolumn{2}{c}{QPO r.m.s \%} &   &  \\
      \hline
    energy band (keV) & additive PDS model & multiplicative PDS model &   &  \\
    \hline
    1.0-2.6 & $9.55\pm1.22$ & $10.04\pm0.58$ &   &  \\
    2.6-4.8 & $9.47\pm1.28$ & $7.76\pm0.79$ &   &  \\
    4.8-7.0 & $8.57\pm1.65$ & $6.63\pm1.30$ &   &  \\
    7.0-11.0 & $9.29\pm1.89$ & $7.57\pm0.98$ &   &  \\
    11.0-23.0 & $4.53\pm1.19$ & $3.35\pm2.41$ &   &  \\
    25.0-35.0 & $8.76\pm1.44$ & $5.93\pm0.92$ &   &  \\
    35.0-48.0 & $5.55\pm2.10$ & $6.16\pm5.68$ &   &  \\
    48.0-67.0 & $9.42\pm0.69$ & $4.51\pm0.55$ &   &  \\
    67.0-100.0 & $6.37\pm0.88$ & $4.04\pm0.77$ &   &  \\
    100.0-150.0 & $5.59\pm1.07$ & $3.54\pm0.71$ &   &  \\
    150.0-200.0 & $4.43\pm0.71$ & $4.77\pm0.35$ &   &  \\
    \hline
    \end{tabular}%
  \label{tab:QPO_harmonic}%
\end{table*}%

\begin{table*}
  \centering
  \caption{The same as table \ref{tab:ACPDSfit} but for the BBN components.}
    \begin{tabular}{ccccc}
    \hline
      & \multicolumn{2}{c}{BBN1 FWHM (Hz)} & \multicolumn{2}{c}{BBN2 FWHM (Hz)} \\
      \hline
    energy band (keV) & additive PDS model & multiplicative PDS model & additive PDS model & multiplicative PDS model \\
    \hline
    1.0-2.6 & $0.48\pm0.08$ & $0.27\pm0.04$ & $3.17\pm0.51$ & $3.01\pm0.35$ \\
    2.6-4.8 & $0.62\pm0.19$ & $0.28\pm0.04$ & $3.89\pm0.76$ & $3.02\pm0.44$ \\
    4.8-7.0 & $0.87\pm0.47$ & $0.30\pm0.06$ & $3.70\pm1.61$ & $2.54\pm0.61$ \\
    7.0-11.0 & $0.54\pm0.24$ & $0.32\pm0.09$ & $10.27\pm9.14$ & $5.56\pm3.56$ \\
    11.0-23.0 & $0.92\pm1.92$ & $0.42\pm0.38$ & $2.56\pm1.59$ & $2.25\pm0.44$ \\
    25.0-35.0 & $0.69\pm0.35$ & $0.38\pm0.07$ & $4.70\pm1.89$ & $3.26\pm0.95$ \\
    35.0-48.0 & $0.58\pm0.80$ & $\cdots$ & $6.24\pm15.27$ & $2.06\pm2.09$ \\
    48.0-67.0 & $0.65\pm0.20$ & $0.41\pm0.05$ & $10.83\pm2.89$ & $2.77\pm0.52$ \\
    67.0-100.0 & $0.69\pm0.54$ & $0.43\pm0.09$ & $2.74\pm0.82$ & $2.44\pm0.35$ \\
    100.0-150.0 & $0.71\pm0.45$ & $0.50\pm0.08$ & $3.76\pm1.71$ & $3.10\pm0.87$ \\
    150.0-200.0 & $1.86\pm0.18$ & $0.61\pm0.08$ & $18.88\pm12.81$ & $6.55\pm2.43$ \\
    \hline
      & \multicolumn{2}{c}{BBN1 r.m.s \%} & \multicolumn{2}{c}{BBN2 r.m.s \%} \\
      \hline
    energy band (keV) & additive PDS model & multiplicative PDS model & additive PDS model & multiplicative PDS model \\
    \hline
    1.0-2.6 & $16.32\pm1.30$ & $\cdots$ & $20.12\pm1.18$ & $22.07\pm0.67$ \\
    2.6-4.8 & $14.36\pm2.21$ & $\cdots$ & $20.56\pm1.19$ & $23.23\pm0.73$ \\
    4.8-7.0 & $14.40\pm5.28$ & $\cdots$ & $18.60\pm3.27$ & $22.31\pm1.29$ \\
    7.0-11.0 & $11.55\pm2.11$ & $\cdots$ & $15.69\pm5.61$ & $15.50\pm2.27$ \\
    11.0-23.0 & $21.02\pm0.75$ & $\cdots$ & $19.22\pm7.50$ & $20.85\pm1.03$ \\
    25.0-35.0 & $9.87\pm2.65$ & $\cdots$ & $14.66\pm1.34$ & $16.88\pm0.92$ \\
    35.0-48.0 & $11.34\pm1.83$ & $\cdots$ & $7.87\pm3.83$ & $11.38\pm4.97$ \\
    48.0-67.0 & $8.01\pm1.10$ & $\cdots$ & $10.53\pm0.93$ & $13.04\pm0.59$ \\
    67.0-100.0 & $6.19\pm4.02$ & $\cdots$ & $13.18\pm1.96$ & $14.56\pm0.56$ \\
    100.0-150.0 & $5.33\pm2.21$ & $\cdots$ & $9.17\pm1.30$ & $10.51\pm0.66$ \\
    150.0-200.0 & $10.49\pm0.56$ & $\cdots$ & $11.46\pm4.92$ & $9.34\pm0.77$ \\
    \hline
    \end{tabular}%
  \label{tab:BBN}%
\end{table*}%

\begin{figure}
    \centering
    \includegraphics[scale=0.5]{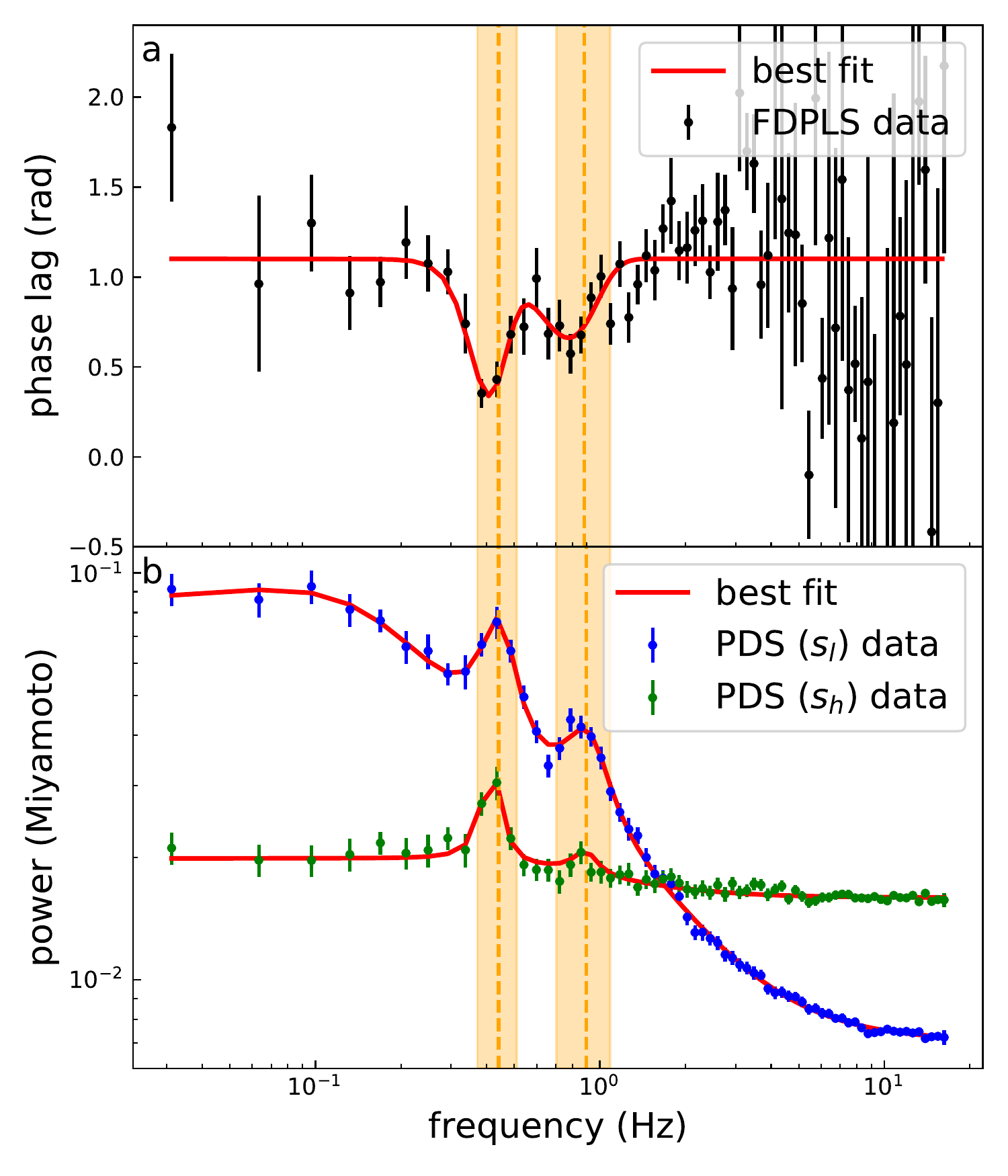}
    \caption{Modeling results for the FDPLS between signals $s_{\rm l}$ and $s_{\rm h}$ and their respective PDS (see section \ref{sec:3} for the definitions of $s_{\rm l}$ and $s_{\rm h}$). In panels a and b, the solid lines are the best-fit using the models proposed in section \ref{sec:3} and the best-fit parameters are listed in table \ref{tab:ACPDSfit}. The yellow bands running through the two panels are the QPO FWHM frequency ranges (including fundamental and harmonic frequencies). Error bars correspond to $1\sigma$ confidence intervals.}
    \label{fig:model_MAXI_FDPLS}
\end{figure}
\begin{table*}
  \caption{Timing properties of $s_{\rm lq}$,$s_{\rm lnq}$,$s_{\rm hq}$ and $s_{\rm hnq}$ (see section \ref{sec:3} for the definition of them). }
    \begin{tabular}{ccccccc}
    \toprule
    signals & bin size (s) & mean rate (cts/s) & exposure (s) & fractional r.m.s & PDS type & FDPLS type \\
    \midrule
    $s_{\rm lq}$ & 0.03125 & 142.5 & 8000 & 27.00\% & QPO & \multirow{2}[0]{*}{dip} \\
    $s_{\rm hq}$ & 0.03125 & 63.5 & 8000 & 9.43\% & QPO & \\
    \hline
    $s_{\rm lnq}$ & 0.03125 & 142.5 & 8000 & 51.00\% & Non-QPO & \multirow{2}[0]{*}{constant} \\
    $s_{\rm hnq}$ & 0.03125 & 63.5 & 8000 & 15.70\% & Non-QPO & \\
    \bottomrule
    \end{tabular}%
  \label{tab:simu_PDS_FDPLS}%
\end{table*}%
\begin{table*}
  \centering
  \caption{PDS and FDPLS fitting results for $s_{\rm l}$ and $s_{\rm h}$.}
    \begin{tabular}{cccccc}
    \toprule
    \multicolumn{2}{c}{$s_{\rm l}$ PDS model} & \multicolumn{2}{c}{$s_{\rm h}$ PDS model} & \multicolumn{2}{c}{FDPLS model} \\
    \midrule
    parameter name & value & parameter name & value & parameter name & value \\
    \hline
    $K_1$ & $0.007\pm0.002$ & $K_1$ & $0.001\pm0.000$ & $A_1$ & $-0.109\pm0.015$ \\
    $f_{c1}$ & $0.442\pm0.008$ & $f_{c1}$ & $0.430\pm0.010$ & $\mu_1$ & $0.405\pm0.011$ \\
    $\omega_1$ & $0.119\pm0.033$ & $\omega_1$ & $0.090\pm0.030$ & $\sigma_1$ & $0.062\pm0.008$ \\
    $K_2$ & $0.009\pm0.002$ & $K_2$ & $0.001\pm0.000$ & $A_2$ & $-0.204\pm0.041$ \\
    $f_{c2}$ & $0.900\pm0.020$ & $f_{c2}$ & $0.860$ (frozen) & $\mu_2$ & $0.776\pm0.037$ \\
    $\omega_2$ & $0.370\pm0.083$ & $\omega_2$ & $0.200$ (frozen) & $\sigma_2$ & $0.185\pm0.024$ \\
    $K_3$ & $0.027\pm0.004$ & $K_3$ & $0.006$ (frozen) & c  & $1.102\pm0.040$ \\
    $f_{c3}$ & $0.000$ (frozen) & $f_{c3}$ & $0.000$ (frozen) & $\cdots$ & $\cdots$ \\
    $\omega_3$ & $0.477\pm0.080$ & $\omega_3$ & $2.061\pm0.425$ & $\cdots$ & $\cdots$ \\
    $K_4$ & $0.040\pm0.005$ & c  & $0.016\pm0.000$ & $\cdots$ & $\cdots$ \\
    $f_{c4}$ & $0.000$ (frozen) & $\cdots$ & $\cdots$ & $\cdots$ & $\cdots$ \\
    $\omega_4$ & $3.175\pm0.512$ & $\cdots$ & $\cdots$ & $\cdots$ & $\cdots$ \\
    c  & $0.007\pm0.000$ & $\cdots$ & $\cdots$ & $\cdots$ & $\cdots$ \\
    \bottomrule
    \end{tabular}%
  \label{tab:fit_PDS_FDPLS}%
\end{table*}%

In addition, we would like to know how the phase lags should look like in MAXI J1820 for additive and convolved signals, so some simulations are done. We use the same data (ObsID P0114661078) used above as an example to show how effective this correction is. From panel h and panel i of Fig.~\ref{fig:MAXI_FDPLS} we can see that the QPO phase lags between the high and low energy bands before and after correction are completely flipped. We focus on two energy bands of these signals: the reference energy band (1-2.6 keV) and the high energy band (100-150 keV). The phase lag between the reference energy band and the high energy band is indicated by the red arrows in panel h and panel i of Fig.~\ref{fig:MAXI_FDPLS}. After the data reduction in the same way as Ma21, we extract two light curves in reference energy band (denote as $s_{\rm l}$) and high energy band (denote as $s_{\rm h}$). The mean count rate of $s_{\rm l}$ and $s_{\rm h}$ are 285 counts/s and 127 counts/s, respectively, and both of the effective exposure time are 8 ks. The FDPLS and PDS of $s_{\rm l}$ and $s_{\rm h}$ are first fitted to obtain the best models, and after that, the best models are used for simulations. For the FDPLS between $s_{\rm l}$ and $s_{\rm h}$, the model takes the form 
\begin{equation}
    {\rm lag}(f) =\frac{A_1}{\sigma_1\sqrt{2\pi}}e^{[{-{(f-\mu_1)^2}/{{2\sigma_1}^2}}]}+\frac{A_2}{\sigma_2\sqrt{2\pi}} e^{[{-{(f-\mu_2)^2}/{{2\sigma_2}^2}}]}+c,
\end{equation}
where the first and second terms represent the dip-like phase lags of the QPO components (including fundamental and harmonic frequencies) and the last term represent the phase lags of the non-QPO components (BBN components). For the PDS of $s_{\rm l}$ and $s_{\rm h}$, a constant term and the sum of four Lorentzian functions are used to fit the data, i.e., the PDS model is the same as equation (\ref{eq:ASM}). We find that the above PDS model does not require the high-frequency zero-centred Lorentzian component for $s_{\rm h}$ when fitting the PDS of $s_{\rm h}$. Moreover, in fitting the PDS of $s_{\rm h}$ we found that the harmonic frequency component of QPO is not well constrained due to the low signal-to-noise ratio of the data. We thus fix the parameters of the QPO harmonic frequency component, which does not affect the goodness of fit, but is useful for our subsequent simulation of the QPO components. The fitting results are shown in Fig. \ref{fig:model_MAXI_FDPLS} and the best-fit parameters of the above models are listed in table \ref{tab:fit_PDS_FDPLS}. It is worth pointing out that we are using an additive PDS model to fit the PDS here, which is correct for additive signal, but not for convolved signal, which should use a multiplicative PDS model. However, we note that the FDPLS relationship of the convolved signal depends only on the FDPLS of the sub-signals and is independent of the PDS of the sub-signals, so the PDS model we use here has no effect on the FDPLS calculation of the convolved signal.

\begin{figure}
    \centering
    \includegraphics[scale=0.36]{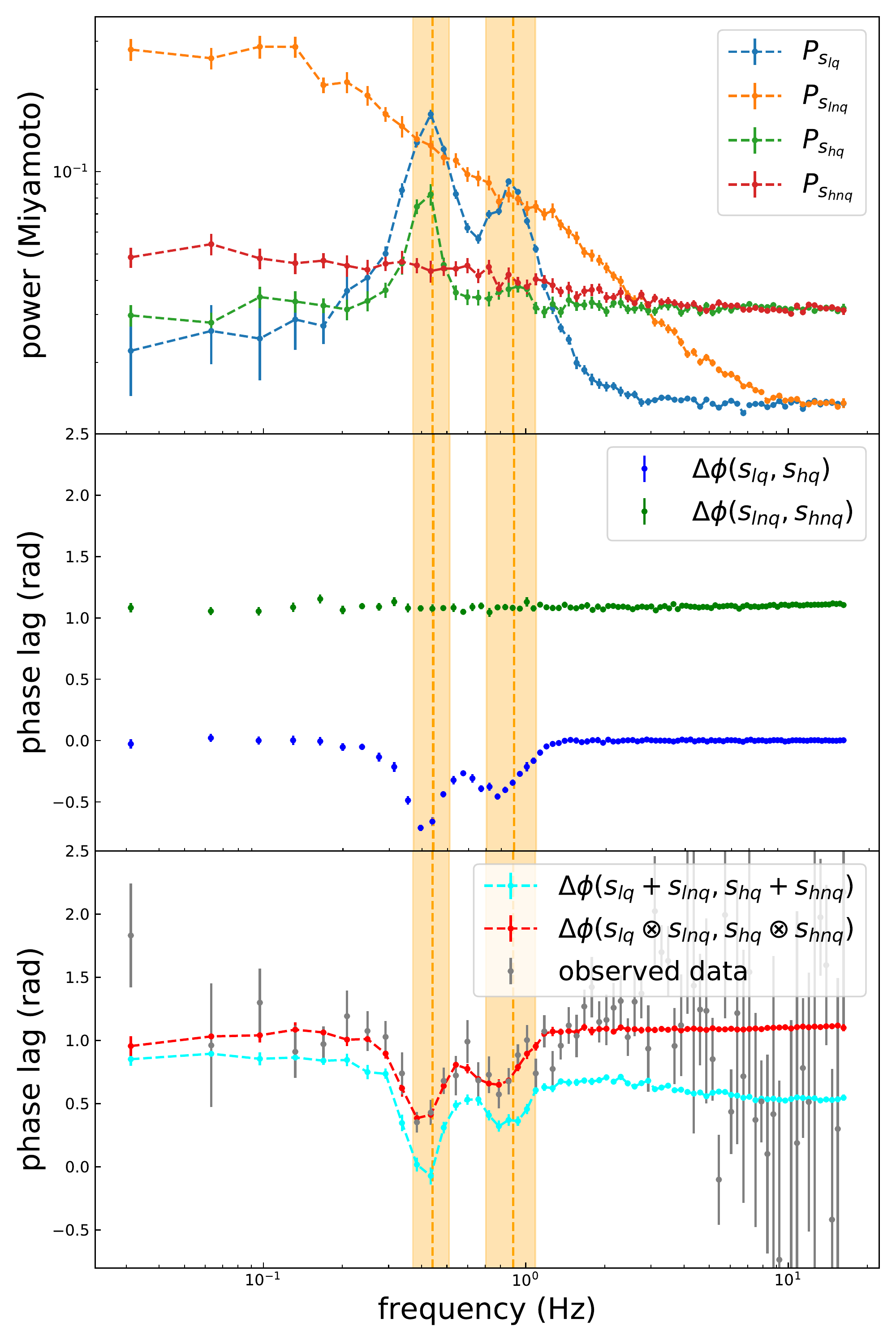}
    \caption{Simulation results based on the data modeling shown in Fig.~\ref{fig:model_MAXI_FDPLS}. Upper panel: the simulated PDS of the signals $s_{\rm lq}$,$s_{\rm lnq}$,$s_{\rm hq}$ and $s_{\rm hnq}$. Middle panel: the simulated FDPLS (blue dots) of $s_{\rm lq}$ and $s_{\rm hq}$ and the simulated FDPLS (green dots) of $s_{\rm lnq}$ and $s_{\rm hnq}$. Bottom panel: the calculated FDPLS of the total signal when the total signal is an/a additive/convolved signal (cyan/red data points). The gray dots are the observed FDPLS. The yellow bands running through panels are the QPO FWHM frequency ranges (including fundamental and harmonic frequencies). Error bars correspond to $1\sigma$ confidence intervals.}
    \label{fig:simu_MAXI_results}
\end{figure}

We then simulate four sub-signals based on the best FDPLS and PDS models obtained above. The PDS of the QPO component is modeled using the sum of the non-zero centred Lorentzian components and the PDS of the BBN component is modeled using the sum of the zero-centred Lorentzian components. We first simulate four signals using TK95 algorithm, noted as $s_{\rm lq}$, $s_{\rm lnq}$, $s_{\rm hq}$, $s_{\rm hnq}$, which stand for the QPO component and the BBN component in the reference energy band, and the QPO component and the BBN component in the high energy band, respectively. After that, we use the algorithm proposed in section \ref{sec:2.3} to make the FDPLS of $s_{\rm lq}$ and $s_{\rm hq}$ satisfy the Gaussian components of the best-fit model and make the FDPLS of $s_{\rm lnq}$ and $s_{\rm hnq}$ satisfy the constant component of the best-fit model. The timing properties of these four signal are listed in table \ref{tab:simu_PDS_FDPLS}. Note that in calculating the FDPLS we split the signal into QPO and non-QPO components, which in effect assumes that the contribution of the additive component of equation (\ref{eq:PDS}) to the overall FDPLS can be neglected, i.e. the destruction of this additive component to the 
additivity of the phase lag between the convolution components can be neglected, as we will explain in detail in the subsequent discussion section. The PDS of $s_{\rm lq}$, $s_{\rm lnq}$, $s_{\rm hq}$, $s_{\rm hnq}$ are shown in top panel of Fig.~\ref{fig:simu_MAXI_results}. The FDPLS between $s_{\rm lq}$ and $s_{\rm hq}$ are two dips, and the FDPLS between $s_{\rm lnq}$ and $s_{\rm hnq}$ are constant, which are shown in the middle panel of Fig.~\ref{fig:simu_MAXI_results}.

Then, $s_{\rm lq}$ and $s_{\rm lnq}$ are added/convolved to get the additive/convolved signal. $s_{\rm hq}$ and $s_{\rm hnq}$ are added/convolved to get the other additive/convolved signal. The FDPLS of the additive/convolved signals is shown by the cyan/red dotted lines in the lower panel of Fig.~\ref{fig:simu_MAXI_results}. The gray dots in the figure are the observed data. As can be seen from Fig.~\ref{fig:simu_MAXI_results}, the simulated results for the additive signal are very different from those given by the data. But the simulated results for the convolved signal match the data perfectly. This simulation result indicates that the observed data can distinguish between convolved and additive signals for FDPLS.
\section{discussion and summary}
\label{sec:4}
In this paper, we investigate the mechanism behind the phase lag correction that was successfully applied for the first time by Ma21 for MAXI J1820+070, where the strong BBN and the QPO coexist. After correcting the phase lag of the QPO, the absolute value of QPO phase lag increases monotonically with photon energy in all observations. Ma21 explained the phase lag behavior of the QPO by employing a compact jet with precession. In this scenario, the high-energy photons come from the part of the jet closer to the black hole, and the precession of the compact jet causes the QPO phenomenon and allows the high-energy photons to reach the observer first, resulting in a soft lag. Because the phase lag behavior can have a large impact on physical conclusions, it is necessary to investigate the rationality of this correction method. Since we want to obtain the intrinsic properties of the QPO, and what we observe is some kind of superposition of the QPO and the BBN components, we have to face the question of how these components constitute the total signal. We found that the correction method is effective only when the sub-signals are synthesized into the total signal by convolution. If the total signal is the convolved signal, the intrinsic phase lags of the QPO can be obtained by subtracting the phase lags of the BBN component from the original phase lags of the total signal, as successfully implemented in Ma21.

If the observed total signals are convolved signals, the corresponding PDS cannot be fitted simply by summing a series of Lorentzian functions (conventions in most of the literatures) but require a multiplicative PDS model. We then try to introduce the convolution mechanism by assuming the propagation of the QPO waves in the corona (may be due to the magneto-acoustic wave propagating within the corona, e.g. \citealt{2010MNRAS.404..738C}). The fluctuation propagation in the form of Dirac delta function resulting in the Green's function. Any form of timing fluctuation will be the convolution of that fluctuation and the Green's function (\citealt{RN958}). We assume that the Green's function is first convolved with the white noise and then convolved with the QPO signal to form the low-frequency part of the observed signal, while the high-frequency part is the result of the convolution of the Green's function with the two white noise 
components. If the Green's function and the QPO signal are convoluted in the time domain, the total PDS will be the multiplication of their respective PDS according to the convolution theorem. Based on this, we introduce a multiplicative PDS model to fit the observed PDS in a representative \emph{Insight}-HXMT observation in 11 different energy bands. For comparison, we also fitted the same data using the traditional additive PDS model. Overall, both additive and multiplicative PDS models fit the observed data well, but the individual components have some differences. The two models give little difference in the centroid frequency as well as in the FWHM of the QPO . For the fundamental frequency component of the QPO, the fraction of r.m.s of the QPO given by the traditional additive PDS model is about $2\sim 3$ higher than that of the multiplicative PDS model, but the trend is the same for both results. For the harmonic frequency component of QPO, the fractional rms given by the two models are not significantly different.

For the traditional additive PDS model, the low-frequency zero-centred Lorentzian component can be considered as the variability due to the propagation of the white noise fluctuation in the outer region of the corona (\citealt{RN970}), and the narrow Lorentzian components stand for the fundamental and harmonic components of the QPO, and the high frequency zero-centred Lorentzian component is responsible for the variability due to the propagation of the white noise fluctuation in the inner region of the corona (\citealt{RN970}). All these terms are simply added together, which means that there is no coherence between them. 

\begin{figure}
    \centering
    \includegraphics[scale=0.5]{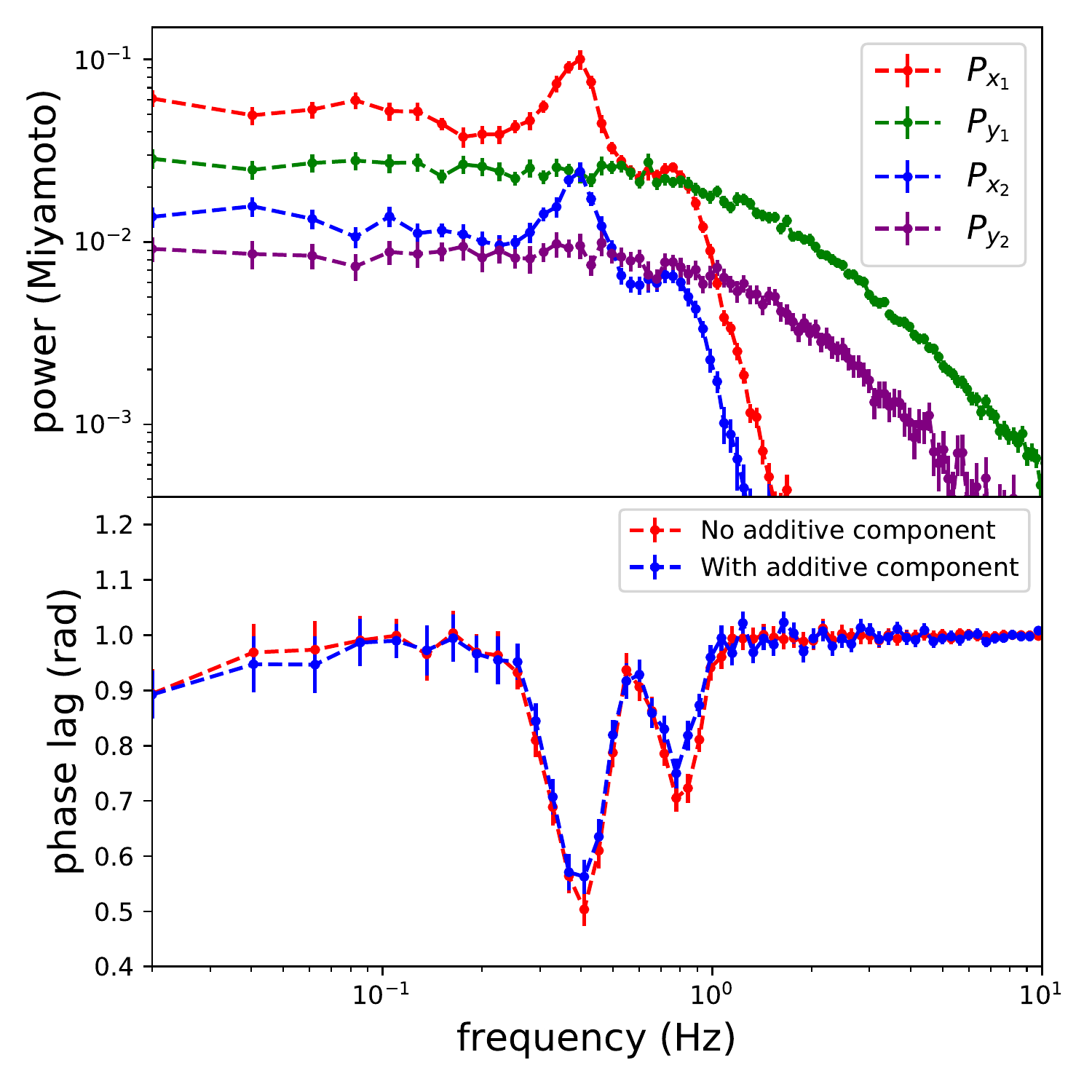}
    \caption{The effect of an additional additive components on the total FDPLS. Upper panel: PDS of simulated signals. Lower panel: FDPLS with or without an additional additive component.}
    \label{fig:with_additive_lag}
\end{figure}

For the multiplicative PDS model, we find that the low frequency component of the PDS can be fitted by multiplying the fundamental and harmonic components of the QPO with a zero-centred Lorentzian function, in addition to an additional additive component to produce the high frequency part of the PDS. The additive Lorentzian component plays the same as the role in the additive PDS model. Therefore, for our multiplicative model, the high-frequency part does not need to be convolved to the QPO signal, but is simply added together. This additive component appearing in the PDS model looks to destroy the additivity of the phase lag brought about by the time domain convolution. We ignored the contribution of this additive component in our previous phase lag calculations. We make a simulation to investigate the effect of this additional additive component on the total FDPLS. Due to the dependence of the FDPLS of additive signals on the PDS of the individual components, we need to know the PDS parameters of each component. Specifically, we first obtain the PDS parameters for each component based on the results of the fit of the multiplicative PDS model (shown in the right panels of Fig. \ref{fig:ACPDS}), and then simulate four signals based on the mean of these parameters in two energy bands: $x_1$ (the convolved signal of energy band 1), $y_1$ (the additional additive signal of energy band 1), $x_2$ (the convolved signal of energy band 2) and $y_2$ (the additional additive signal of energy band 2). The PDS of these four signals are shown in the upper panel of Fig. \ref{fig:with_additive_lag}. We then set the FDPLS model of $x_1$ and $x_2$ to $\phi(f)=-0.5e^{-\frac{(f-0.4)^2}{0.1^2}}-0.3e^{-\frac{(f-0.8)^2}{0.15^2}}+1$ and set the FDPLS model of $y_1$ and $y_2$ to 1. Finally we calculate the FDPLS of $x_1$ and $x_2$ and the FDPLS of $x_1+y1$ and $x_2+y_2$, respectively. By comparing these two FDPLS, the effect of the additional additive component on the total FDPLS can be known. As can be seen from the lower panel of Fig. \ref{fig:with_additive_lag}, the additional additive component has almost no effect on the total FDPLS, except for a slight dilution of the phase lag of the harmonic component. It is therefore reasonable to ignore the effect of the additional additive component on the FDPLS of the convolution components in our previous analysis.

Traditionally, it is mostly assumed that the observed components are additive in the time domain, and it has also been suggested that it might be more reasonable to multiply these components in the time domain based on the fluctuation propagation model (e.g. \cite{RN966}). However, neither of these two models can explain the phase lag correction in Ma21. In order to explain the correction of phase lags in Ma21, we propose a convolution model instead of the additive and multiplicative models in the time domain, which is supported by comparison between simulations and data on both PDS and FDPLS. This suggests that the convolution model can explain the behaviour of the phase lag observed in MAXI J1820+070, in which case the phase lag correction method applied in Ma21 is correct. 

Finally, it is worth pointing out that our current convolution model still has limitations. For examples, it is not yet possible to explain the energy dependence of the phase lag using the convolution model, and the relationship of individual components to specific physical processes needs further development. However, it is certain that at least part of the time domain signal is filtered by the system before it reaches the observer (e.g. both the accretion disk and corona/jet play the role of low-pass filters to some extent), and these response processes are necessarily accompanied by time domain convolution operations.

\section*{Acknowledgments}
This research has made use of the data from the \emph{Insight}-HXMT mission, a project funded by China National Space Administration and the Chinese Academy of Sciences. This work is supported by the National Natural Science Foundation of China under
grants 12133007, U1838201, U1938201.
\section*{Data Availability}
The data used in this paper can be found in the \emph{Insight}-HXMT website (\url{http://hxmtweb.ihep.ac.cn/}).


\bibliographystyle{mnras}
\bibliography{mnras.bib} 

\bsp	
\label{lastpage}
\end{document}